\documentclass[a4paper,fleqn,reqno]{amsart}
\usepackage{epic,eepic}
\usepackage{epsfig,cite,indentfirst}
\usepackage{multicol,boxedminipage}
\usepackage{varioref} 
\begin{document}

\title[Factorized vertex DWPF]
{Factorized domain wall partition functions 
 in trigonometric vertex models} 

\author{O Foda, M Wheeler and M Zuparic}

\address{Department of Mathematics and Statistics,
         University of Melbourne, 
         Parkville, Victoria 3010, Australia.}

\email{foda, mwheeler, mzup@ms.unimelb.edu.au}

\keywords{Vertex models. 
          Domain wall boundary conditions} 
\subjclass[2000]{Primary 82B20, 82B23}
\date{}

\newcommand{\field}[1]{\mathbb{#1}}
\newcommand{\CC}{\field{C}}
\newcommand{\NN}{\field{N}}
\newcommand{\ZZ}{\field{Z}}

\newcommand{\B}{{\mathcal B}}
\newcommand{\D}{{\mathcal D}}
\newcommand{\M}{{\mathcal M}}
\newcommand{\R}{{\mathcal R}}
\newcommand{\T}{{\mathcal T}}
\newcommand{\U}{{\mathcal U}}
\newcommand{\Y}{{\mathcal Y}}
\newcommand{\Z}{{\mathcal Z}}

\newcommand{\tT}{\widetilde{\mathcal T}}
\newcommand{\tU}{\widetilde{\mathcal U}}

\renewcommand{\P}{{\mathcal P}}

\begin{abstract}
We obtain factorized domain wall partition functions for two sets of 
trigonometric vertex models: 
{\bf 1.} The $N$-state Deguchi-Akutsu models, for $N \in \{2, 3, 4\}$ 
(and conjecture the result for all $N \geq 5$), and
{\bf 2.} The $sl(r + 1| s + 1)$ Perk-Schultz models, for $\{r, s \in 
\NN\}$, where (given the symmetries of these models) the result is 
independent of $\{r, s\}$.
\end{abstract}

\maketitle

\newtheorem{ca}{Figure}
\newtheorem{corollary}{Corollary}
\newtheorem{definition}{Definition}
\newtheorem{example}{Example}
\newtheorem{lemma}{Lemma}
\newtheorem{notation}{Notation}
\newtheorem{proposition}{Proposition}
\newtheorem{remark}{Remark}
\newtheorem{theorem}{Theorem}
\newtheorem{property}{Property}

\def\ll{\left\lgroup}
\def\rr{\right\rgroup}

\newcommand{\Proof}{\medskip\noindent {\it Proof: }}
\def\proofend{\ensuremath{\square}}

\def\no{\nonumber}

\def\union{\mathop{\bigcup}}
\def\vac{|\mbox{vac}\rangle}
\def\cav{\langle\mbox{vac}|}

\def\lprod{\mathop{\prod{\mkern-29.5mu}{\mathbf\longleftarrow}}}
\def\rprod{\mathop{\prod{\mkern-28.0mu}{\mathbf\longrightarrow}}}

\def\r{\rangle}
\def\l{\langle}

\def\a{\alpha}
\def\b{\beta}
\def\g{\gamma}
\def\d{\delta}
\def\e{\epsilon}
\def\l{\lambda}

\def\s{\sigma}

\def\eps{\varepsilon}
\def\hb{\hat\beta}

\def\tg{\operatorname{tg}}
\def\ctg{\operatorname{ctg}}
\def\sh{\operatorname{sh}}
\def\ch{\operatorname{ch}}
\def\cth{\operatorname{cth}}
\def\th{\operatorname{th}}

\def\tla{\tilde{\lambda}}
\def\tmu{\tilde{\mu}}

\def\sul{\sum\limits}
\def\pl{\prod\limits}

\def\pd #1{\frac{\partial}{\partial #1}}
\def\const{{\rm const}}
\def\argum{\{\mu_j\},\{\la_k\}} 
\def\umarg{\{\la_k\},\{\mu_j\}} 

\def\prodmu #1{\prod\limits_{j #1 k} \sinh(\mu_k-\mu_j)}
\def\prodla #1{\prod\limits_{j #1 k} \sinh(\lambda_k-\lambda_j)}

\newcommand{\bl}[1]{\makebox[#1em]{}}

\def\tr{\operatorname{tr}}
\def\Res{\operatorname{Res}}
\def\det{\operatorname{det}}

\newcommand{\boldN}{\boldsymbol{N}}
\newcommand{\bra}[1]{\langle\,#1\,|}
\newcommand{\ket}[1]{|\,#1\,\rangle}
\newcommand{\bracket}[1]{\langle\,#1\,\rangle}
\newcommand{\infinity}{\infty}

\renewcommand{\labelenumi}{\S\theenumi.}

\let\up=\uparrow
\let\down=\downarrow
\let\tend=\rightarrow
\hyphenation{boson-ic
             ferm-ion-ic
             para-ferm-ion-ic
             two-dim-ension-al
             two-dim-ension-al
             rep-resent-ative
             par-tition}

\renewcommand{\mod}{\textup{mod}\,}
\newcommand{\wt}{\text{wt}\,}

\hyphenation{And-rews
             Gor-don
             boson-ic
             ferm-ion-ic
             para-ferm-ion-ic
             two-dim-ension-al
             two-dim-ension-al}

\setcounter{section}{-1}

\section{Introduction}

{\bf Domain wall partition functions (DWPF's)} were first proposed 
and evaluated in determinant form, for the spin-$\frac{1}{2}$ vertex 
model\footnote{In the sequel, {\sl `model'} will always mean 
{\sl `trigonometric vertex model'}.} on a finite square lattice, 
in \cite{korepin, izergin}. At the free fermion point of the 
spin-$\frac{1}{2}$ model, this determinant is in Cauchy form and 
therefore factorizes. More recently, determinant expressions for 
the DWPF's of spin-$\frac{N-1}{2}$ models and also of level-1 affine 
$so(N)$ models (for certain discrete values of the crossing parameter) 
were obtained in \cite{cfk} and in \cite{df}, respectively.  

{\bf State variable conjugation.} We are interested in models 
with state variables $\{\sigma\}$. Each state variable takes discrete 
integral values, $\sigma \in \{1, \cdots, N\}$. We define {\sl `state 
variable conjugation'} as replacing each state variable 
$\sigma$ by $(N - \sigma + 1)$. The models mentioned above are 
invariant under this conjugation. 

{\bf The $N$-state Deguchi-Akutsu ($N$--DA) models}, $N \geq 2$ 
\cite{da1}, are models with vertex weights\footnote{In the sequel, 
{\sl `weight'} will always stand for {\sl `vertex weight'}.}, that 
depend on two sets of parameters: 
{\bf 1.} Vertical and horizontal rapidities, and 
{\bf 2.} Vertical and horizontal external field variables.
They reduce in the limit of no external fields to the 
spin-$\frac{N-1}{2}$ models at their respective 
free fermion points. 

{\bf Factorized DWPF.} In \cite{4-authors}, a determinant expression 
for the DWPF of the 2--DA model was obtained using the arguments of 
\cite{korepin, izergin}, but only for zero values of the rapidities. 
This determinant is in Cauchy form and therefore factorizes. For 
general values of all parameters, no determinant expression was
found, and it was argued on general grounds that no such expression 
exists. However, using the {\it F}--basis of \cite{kmt}, a factorized 
expression for the 2--DA DWPF was obtained. 

{\bf In this work,} we extend the above result to the $N$-DA models, 
$N \in \{2, 3, 4\}$. Our results are restricted to $N \in \{2, 3, 4\}$ 
because our proofs require the explicit expressions of the weights, 
while the number of vertices grows $\sim O(N^3)$. However, our results 
are quite simple and have a uniform dependence on $N$, which allows 
us to conjecture that our expression extends to all $N \geq 2$. 

{\bf Non-invariance under state variable conjugation.} Our proofs rely 
on the non-invariance of the $N$-DA models under state variable 
conjugation (for non-vanishing external fields). This leads us to look 
for other models that are similarly non-invariant. 

{\bf The $sl(r+1|s+1)$ Perk-Schultz (PS) models}, $\{r, s \in \NN\}$ 
\cite{ps}, form another class of models that are non-invariant under 
state variable conjugation, in this case because the state variables 
belong to two different sets with different statistics. 

In these models, the definition of domain wall boundary conditions 
(DWBC's) is not unique. From experience with the $N$-DA models, we 
propose a definition that leads to factorized DWPF's. The symmetries 
of the Perk-Schultz models are such that the result is independent of 
$\{r, s\}$. 

{\bf Outline of paper.} 
In section {\bf 1}, we recall basic definitions, introduce the 
$N$--DA models and obtain the corresponding factorized DWPF. 
In section {\bf 2}, we do the same for the $sl(r+1|s+1)$ 
Perk-Schultz models. Section {\bf 3} contains brief remarks and 
appendix {\bf A} lists the weights of the $N$--DA models, for 
$N \in \{2, 3, 4\}$.

\section{The $N$-state Deguchi-Akutsu ($N$--DA) models}\label{DA}

\begin{multicols}{2}

\subsection{The lattice} We work on a square lattice consisting 
of $L$ vertical and $L$ horizontal lines, label the vertical lines 
from left to right and the horizontal from top to bottom.

We assign the $i$-th vertical line an orientation from bottom 
to top, a complex rapidity variable $u_i$ and a complex external
field variable $\alpha_i$. We assign the $j$-th horizontal line 
an orientation from left to right, a complex rapidity variable 
$v_j$ and a complex external field variable $\beta_j$.

%
\begin{center}
\begin{minipage}{2.2in}
\setlength{\unitlength}{0.0008cm}
\begin{picture}(4800, 6000)(-1000, -0300)
\thicklines
\path(2400,5400)(2400,1800)
\path(3000,5400)(3000,1800)
\path(3600,5400)(3600,1800)
\path(4200,5400)(4200,1800)
\path(4800,5400)(4800,1800)
\path(1800,4800)(5400,4800)
\path(1800,4200)(5400,4200)
\path(1800,3600)(5400,3600)
\path(1800,3000)(5400,3000)
\path(1800,2400)(5400,2400)
\path(0600,4254)(1200,4254)
\path(2400,654)(2400,1254)
\path(3000,654)(3000,1254)
\path(3600,654)(3600,1254)
\path(4200,654)(4200,1254)
\path(4800,654)(4800,1254)
\path(600,2454)(1200,2454)
\path(600,3054)(1200,3054)
\path(600,3654)(1200,3654)
\path(600,4854)(1200,4854)
\whiten\path(2490,894)(2400,1254)(2310,894)(2490,894)
\whiten\path(3090,894)(3000,1254)(2910,894)(3090,894)
\whiten\path(3690,894)(3600,1254)(3510,894)(3690,894)
\whiten\path(4290,894)(4200,1254)(4110,894)(4290,894)
\whiten\path(4890,894)(4800,1254)(4710,894)(4890,894)
\whiten\path(840,2364)(1200,2454)(840,2544)(840,2364)
\whiten\path(840,2964)(1200,3054)(840,3144)(840,2964)
\whiten\path(840,3564)(1200,3654)(840,3744)(840,3564)
\whiten\path(840,4164)(1200,4254)(840,4344)(840,4164)
\whiten\path(840,4764)(1200,4854)(840,4944)(840,4764)
%

\put(-0800,4854){$v_1, \beta_1$}
%
\put(-0800,2454){$v_L, \beta_L$}
\put(2300, 0250){$u_1$}
\put(2300,-0250){$\alpha_1$}
%
\put(4700, 0250){$u_L$}
\put(4700,-0250){$\alpha_L$}
\end{picture}
\begin{ca}
\label{lattice}
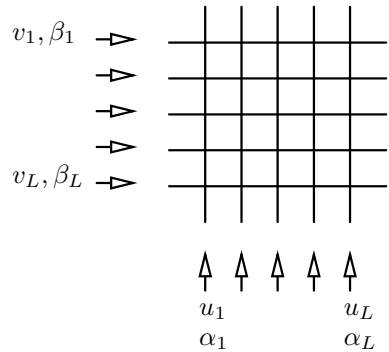
An $L\times L$ square lattice, with oriented lines
and variables.
\end{ca}

\end{minipage}
\end{center}

  \end{multicols}
\begin{multicols}{2}

\subsection{Vertices} Each lattice line intersects with $L$ other
lines. A line segment between two intersections is a bond. To each
bond, we assign a state variable $\sigma \in \{1, 2, \cdots, N\}$.
The intersection of the $i$-th vertical line and the $j$-th 
horizontal line, together with the four bonds adjacent to it, and
the set of state variables on these bonds, is a vertex $v_{ij}$.

%
\begin{center}
\begin{minipage}{2.2in}
\setlength{\unitlength}{0.0008cm}
%


\begin{picture}(4000, 5000)(-1920,-1000)

\thicklines


\path(0620,2220)(3420,2220)
\put(-1400,2120){$v, \beta$}
\whiten\path(-160,2130)(0200,2220)(-160,2310)(-160,2130)
\path(-520,2220)(-160,2220)
\put(0620,2420){$\kappa_2$}
\put(3220,2420){$\iota_2$}


\path(2020,0820)(2020,3620)
\whiten\path(2110,40)(2020,400)(1930,40)(2110,40)
\path(2020,-320)(2020,40)
\put(1920,-720){$u, \alpha$}
\put(2220,1020){$\kappa_1$}
\put(2220,3420){$\iota_1$}
\end{picture}
\begin{ca}
The vertex corresponding to
$X_{\alpha,\beta}(u-v)^{\iota_1,\iota_2}_{\kappa_2,\kappa_1}$.
\label{general-vertex}
\end{ca}
\end{minipage}
\end{center}

\end{multicols}

\bigskip

\subsection{Weights} To each vertex $v_{ij}$ we assign a weight
$w_{ij}$, that depends on the state variables on the four
bonds of that vertex, the difference of rapidity variables
flowing through the vertex, and the two external field
variables flowing through the vertex.
Specifically, a vertex with vertical rapidity and external field
variable $\{u, \alpha\}$, horizontal rapidity and external field
variable $\{v, \beta\}$, and state variables
$\{\iota_1, \iota_2, \kappa_1, \kappa_2\}$ is assigned the weight
$X_{\alpha,\beta}(u-v)^{\iota_1,\iota_2}_{\kappa_2,\kappa_1}$.
These weights satisfy the Yang-Baxter equation:

\begin{eqnarray}
\sum_{\lambda_1, \lambda_2, \lambda_3}
X_{\alpha,\beta}(u-v)^{\iota_1, \iota_2}_{\lambda_2, \lambda_1}
X_{\alpha,\gamma}(u-w)^{\lambda_1, \iota_3}_{\lambda_3,  \kappa_1}
X_{\beta,\gamma}(v-w)^{\lambda_2, \lambda_3}_{\kappa_3, \kappa_2}
=
\\
\sum_{\lambda_1, \lambda_2, \lambda_3}
X_{\beta,\gamma}(v-w)^{\iota_2, \iota_3}_{\lambda_3,  \lambda_2}
X_{\alpha,\gamma}(u-w)^{\iota_1, \lambda_3}_{\kappa_3, \lambda_1}
X_{\alpha,\beta}(u-v)^{\lambda_1,\lambda_2}_{\kappa_2, \kappa_1}
\no
\end{eqnarray}

Expressions for all
$X_{\alpha,\beta}(u-v)^{\iota_1,\iota_2}_{\kappa_2,\kappa_1}$ are given 
in \cite{da1}, for $N \in \{2,3,4\}$. For completeness we include them 
in appendix {\bf A}. From these expressions, one can check that the 
weights of the $N$--DA model are not invariant under conjugating 
the state variables. This is due to the presence of external fields 
$\{\alpha, \beta\}$.

Switching off the external fields restores the symmetry of the
weights (up to global gauge transformations). The (symmetrized)
no-field weights coincide with the weights of the
spin-$\frac{N-1}{2}$ models at their free fermion point. This can be 
checked trivially for $N = 2$ by setting $\alpha = \beta = \sqrt{-1}$.

\begin{multicols}{2}

\subsection{Minimal and maximal state variables} 
We refer to the state variable $\sigma = 1$ as minimal, 
and to $\sigma = N$ as maximal. 


\subsection{The $c_{+}$ vertex} We refer to the unique vertex with 
minimal state variables incoming from the left and exiting from
above, maximal state variables incoming from below and exiting
from the right, as shown in Figure {\bf \ref{c+vertex}}, as the 
$c_{+}$ vertex. In the $N$--DA models
$
c_{+}(\alpha, \beta, u-v) = X_{\alpha,\beta}(u-v)^{1,N}_{1,N}
$.

%
\begin{center}
\begin{minipage}{2.4in}
\setlength{\unitlength}{0.0008cm}
\begin{picture}(4000, 5000)(-1920,-1000)
\thicklines


\path(0620,2220)(3420,2220)
\put(-1520,2120){$v, \beta$}
\whiten\path(-160,2130)(0200,2220)(-160,2310)(-160,2130)
\path(-520,2220)(-160,2220)
\put(0620,2420){$1$}
\put(3220,2420){$N$}


\path(2020,0820)(2020,3620)
\whiten\path(2110,40)(2020,400)(1930,40)(2110,40)
\path(2020,-320)(2020,40)
\put(1920,-720){$u, \alpha$}
\put(2220,1020){$N$}
\put(2220,3420){$1$}
\end{picture}
\begin{ca}
The $c_{+}$ vertex.
\label{c+vertex}
\end{ca}
\end{minipage}
\end{center}

\end{multicols}

\begin{multicols}{2}

\subsection{Domain wall boundary conditions (DWBC)}

We define DWBC's in the $N$--DA model as a form of {\sl expanded
$c_{+}$ vertex}: All (boundary) bonds on the left and top carry 
minimal state variables, while all bonds on the right and 
below carry maximal state variables.

%
\begin{center}
\begin{minipage}{2.4in}
\setlength{\unitlength}{0.0008cm}
\begin{picture}(4800, 5500)(0,-0000)
\thicklines


\path(2400,5400)(2400,1800)
 \put(2450,5400){$1$}
 \put(2400,1600){$N$}

\path(3000,5400)(3000,1800)
 \put(3050,5400){$1$}
 \put(3000,1600){$N$}

\path(3600,5400)(3600,1800)
 \put(3650,5400){$1$}
 \put(3600,1600){$N$}

\path(4200,5400)(4200,1800)
 \put(4250,5400){$1$}
 \put(4200,1600){$N$}

\path(4800,5400)(4800,1800)
 \put(4850,5400){$1$}
 \put(4800,1600){$N$}


\path(1800,4800)(5400,4800)
 \put(1700,4900){$1$}
 \put(5400,4900){$N$}

\path(1800,4200)(5400,4200)
 \put(1700,4300){$1$}
 \put(5400,4300){$N$}

\path(1800,3600)(5400,3600)
 \put(1700,3700){$1$}
 \put(5400,3700){$N$}

\path(1800,3000)(5400,3000)
 \put(1700,3100){$1$}
 \put(5400,3100){$N$}

\path(1800,2400)(5400,2400)
 \put(1700,2500){$1$}
 \put(5400,2500){$N$}

\path(0600,4254)(1200,4254)
\path(2400,0654)(2400,1254)
\path(3000,0654)(3000,1254)
\path(3600,0654)(3600,1254)
\path(4200,0654)(4200,1254)
\path(4800,0654)(4800,1254)
\path(0600,2454)(1200,2454)
\path(0600,3054)(1200,3054)
\path(0600,3654)(1200,3654)
\path(0600,4854)(1200,4854)
\whiten\path(2490,894)(2400,1254)(2310,894)(2490,894)
\whiten\path(3090,894)(3000,1254)(2910,894)(3090,894)
\whiten\path(3690,894)(3600,1254)(3510,894)(3690,894)
\whiten\path(4290,894)(4200,1254)(4110,894)(4290,894)
\whiten\path(4890,894)(4800,1254)(4710,894)(4890,894)
\whiten\path(840,2364)(1200,2454)(840,2544)(840,2364)
\whiten\path(840,2964)(1200,3054)(840,3144)(840,2964)
\whiten\path(840,3564)(1200,3654)(840,3744)(840,3564)
\whiten\path(840,4164)(1200,4254)(840,4344)(840,4164)
\whiten\path(840,4764)(1200,4854)(840,4944)(840,4764)
\put(-100,4854){$$}
\put(-100,3054){$$}
\put(-100,2454){$$}
\put(2300,0250){$$}
\put(2900,0250){$$}
\put(4700,0250){$$}
\end{picture}

\begin{ca}
\label{dwbc}
$N$--DA DWBC's. 
\end{ca}

\end{minipage}
\end{center}

\end{multicols}

\bigskip

\subsection{Line-permuting vertices.} Any model, on a finite 
lattice with DWBC's, has a pair of vertices that can be used 
to permute adjacent lattice lines, as we will see in detail 
below. We call these vertices $\{a_{+}, a_{-}\}$. In the 
$N$--DA models, 

\begin{equation}
\begin{split}
&a_{+}(\alpha, \beta, u-v) = X_{\alpha,\beta}(u-v)^{1,1}_{1,1}\\
&a_{-}(\alpha, \beta, u-v) = X_{\alpha,\beta}(u-v)^{N,N}_{N,N} 
\end{split}
\end{equation}

\subsection{Domain wall partition function (DWPF)} As always,
the DWPF on an $L\times L$ lattice, $Z^{DA}_{L\times L}$, is
the sum over all weighted configurations that satisfy the DWBC.
The weight of each configuration is the product of the weights
of the vertices:

\begin{equation}
Z^{DA}_{L\times L}
=
\sul_{{\rm configurations}}
\ll
\pl_{\rm vertices} w_{ij}
\rr
\label{physical}
\end{equation}

\subsection{Properties of the $N$--DA DWPF} The $N$--DA DWPF,
$N \in \{2, 3, 4\}$, satisfies four properties:

\bigskip
\noindent {\bf Property 1.} From the DWBC's, $Z^{DA}_{L\times L}$ 
has the form

\begin{equation}
Z^{DA}_{L\times L}\ll \{\alpha\},\{\beta\},\{u\},\{v\}\rr
=
e^{(N-1)u_1}p \ll \{\alpha\},\{\beta\},\{u\},\{v\}\rr
\end{equation}

\noindent where $p$ is a polynomial of degree $(L-1)(N-1)$ in 
$e^{u_1}$.

This can be seen from the fact that the rapidity $u_1$ only appears 
in the left-most column, and every vertex in that column is of the 
form $X_{\alpha_1, \beta_j}(u_1-v_j)^{\iota,\kappa}_{1,\lambda}$. 
From appendix {\bf A}, one can check that

\begin{equation}
X_{\alpha_1, \beta_j}(u_1-v_j)^{\iota, \kappa}_{1, \lambda}
=
e^{(\lambda-\iota)u_1}
q(\alpha_1, \beta_j, u_1, v_j)^{\iota, \kappa}_{1, \lambda}
\end{equation}

\noindent where $q^{\iota,\kappa}_{1,\lambda}$ is a polynomial of 
degree $(N-1+\iota-\lambda)$ in $e^{u_1}$. Property {\bf 1} then 
follows by noticing that every lattice configuration in the DWPF 
receives a contribution of 

\begin{equation}
X_{\alpha_1,\beta_1}(u_1-v_1)^{1,\kappa_1}_{1,\lambda_1}
X_{\alpha_1,\beta_2}(u_1-v_2)^{\lambda_1,\kappa_2}_{1,\lambda_2}
\ldots
X_{\alpha_1,\beta_L}(u_1-v_L)^{\lambda_{L-1},\kappa_L}_{1,N}
\end{equation}

\noindent from the left-most column.

\bigskip


\noindent {\bf Property 2.} $Z^{DA}_{L\times L} $ has zeros in 
$e^{u_1}$ at the $(L-1)(N-1)$ points

\begin{equation}
e^{u_1} = \frac{e^{u_k}}{\rho^{j-1} \alpha_1 \alpha_k}, 
\quad
j \in \{1, \ldots, N-1\}, k \in \{2, \ldots, L  \}
\label{zeros}
\end{equation}


\begin{center}
\begin{minipage}{2.4in}
\setlength{\unitlength}{0.0008cm}
\begin{picture}(4800, 5500)(0,0500)
\thicklines


\path(2400,5400)(2400,1800)
 \put(2500,5400){$1$}
 \put(1900,1600){$N$}
 \put(1900,1000){$N$}

\path(3000,5400)(3000,1800)
 \put(3100,5400){$1$}
 \put(3100,1600){$N$}
 \put(3100,1000){$N$}

\path(2400,1800)(3000,1200)
\path(3000,1800)(2400,1200)

\path(3600,5400)(3600,1800)
 \put(3700,5400){$1$}
 \put(3600,1600){$N$}

\path(4200,5400)(4200,1800)
 \put(4300,5400){$1$}
 \put(4200,1600){$N$}

\path(4800,5400)(4800,1800)
 \put(4900,5400){$1$}
 \put(4800,1600){$N$}


\path(1800,4800)(5400,4800)
 \put(1700,4900){$1$}
 \put(5400,4900){$N$}

\path(1800,4200)(5400,4200)
 \put(1700,4300){$1$}
 \put(5400,4300){$N$}

\path(1800,3600)(5400,3600)
 \put(1700,3700){$1$}
 \put(5400,3700){$N$}

\path(1800,3000)(5400,3000)
 \put(1700,3100){$1$}
 \put(5400,3100){$N$}

\path(1800,2400)(5400,2400)
 \put(1700,2500){$1$}
 \put(5400,2500){$N$}

\end{picture}

\begin{ca}
Attaching an $a_{-}$ vertex.
\end{ca}

\end{minipage}
\end{center}


This can be seen as follows: Multiply 
$Z^{DA}_{L\times L} \ll \{\alpha\},\{\beta\},\{u\},\{v\}\rr$
with the vertex $a_{-}(\alpha_1,\alpha_2,u_1-u_2)$. 
This corresponds to attaching a type $a_{-}$ vertex from below, 
as in Figure {\bf 5}.


Using the Yang-Baxter equation, we slide the inserted vertex through 
the lattice until it emerges from the top, as in Figure {\bf 6}.

The inserted
$a_{-}(\alpha_1,\alpha_2,u_1-u_2)$ vertex emerges as an
$a_{+}(\alpha_1,\alpha_2,u_1-u_2)$ vertex, and in the process, 
the two left-most vertical lattice lines are permuted. 


\begin{center}
\begin{minipage}{2.4in}
\setlength{\unitlength}{0.0008cm}
\begin{picture}(4800, 5500)(0,1000)
\thicklines


\path(2400,5400)(3000,6000)
\path(3000,5400)(2400,6000)

\path(2400,5400)(2400,1800)
 \put(2050,5400){$1$}
 \put(2050,6000){$1$}
 \put(2400,1600){$N$}

\path(3000,5400)(3000,1800)
 \put(3050,5400){$1$}
 \put(3050,6000){$1$}
 \put(3000,1600){$N$}

\path(3600,5400)(3600,1800)
 \put(3650,5400){$1$}
 \put(3600,1600){$N$}

\path(4200,5400)(4200,1800)
 \put(4250,5400){$1$}
 \put(4200,1600){$N$}

\path(4800,5400)(4800,1800)
 \put(4850,5400){$1$}
 \put(4800,1600){$N$}


\path(1800,4800)(5400,4800)
 \put(1700,4900){$1$}
 \put(5400,4900){$N$}

\path(1800,4200)(5400,4200)
 \put(1700,4300){$1$}
 \put(5400,4300){$N$}

\path(1800,3600)(5400,3600)
 \put(1700,3700){$1$}
 \put(5400,3700){$N$}

\path(1800,3000)(5400,3000)
 \put(1700,3100){$1$}
 \put(5400,3100){$N$}

\path(1800,2400)(5400,2400)
 \put(1700,2500){$1$}
 \put(5400,2500){$N$}

\end{picture}

\begin{ca}
Extracting an $a_{+}$ vertex.
\end{ca}

\end{minipage}
\end{center}


We conclude that

\begin{eqnarray}
Z^{DA}_{L\times L} \ll \{\alpha\},\{\beta\},\{u\},\{v\}\rr
=
\frac{
a_{+}(\alpha_1, \alpha_2, u_1- u_2)
}
{
a_{-}(\alpha_1, \alpha_2, u_1- u_2)
}
\times
\\
\phantom{\times}
Z^{DA}_{L\times L} \ll \{\alpha_2, \alpha_1, \ldots\},
                   \{\beta\},\{u_2, u_1, \ldots \},
                   \{v\}\rr
\no
\end{eqnarray}

Iterating the above procedure ($L-1$) times, we obtain

\begin{equation}
\begin{split}
Z^{DA}_{L\times L} &\ll \{\alpha\},\{\beta\},\{u\},\{v\}\rr
=
\prod_{j=2}^{L}
\ll
\frac{
a_{+}(\alpha_1,\alpha_j,u_1-u_j)
}
{
a_{-}(\alpha_1,\alpha_j,u_1-u_j)
}
\rr
\times
\\
&Z^{DA}_{L\times L}
\ll
\{ \alpha_2, \ldots, \alpha_L, \alpha_1\}, \{\beta\},
 \{u_2,       \ldots, u_L, u_1 \},\{v\}
\rr
\end{split}
\label{locations}
\end{equation}

The locations of the $(L-1)(N-1)$ zeros in $e^{u_1}$ follow
from Equation {\bf \ref{locations}}. 

\bigskip
\noindent {\bf Property 3.} $Z^{DA}_{L\times L}$ obeys the 
recursion relation

\begin{equation}
\begin{split}
& Z^{DA}_{L\times L}
\left|_{
e^{u_1}=
\frac{\beta_L}{\alpha_1}
e^{v_L}
}
\right.
=
\ll
\frac{\beta_L}{\alpha_1}
\rr^{N-1}
\prod_{j=1}^{N-1}
\ll
\sqrt{1 - \rho^{j-1} \alpha_1^2}
\sqrt{1 - \rho^{j-1}  \beta_L^2}
\rr
\times \\
& \prod_{j=1}^{N-1}
  \prod_{k=1}^{L-1}
\ll
1-\rho^{j-1}\beta_L\beta_k e^{v_L-v_k}
\rr
\prod_{k=2}^{L}
\ll
e^{u_k-v_L}-\rho^{j-1}\alpha_k\beta_L
\rr
Z^{DA, (1L)}_{(L-1)\times (L-1)}
\label{recursion2}
\end{split}
\end{equation}

\noindent where $Z^{DA, (1L)}_{(L-1)\times (L-1)}$ is the DWPF on
an $(L-1)\times (L-1)$ lattice, with the omission of external field
variables $\{\alpha_1, \beta_L \}$ and rapidities $\{u_1, v_L\}$.
This is seen by noting the lower-left vertex must be 
$X_{\alpha_1,\beta_L}(u_1-v_L)^{\iota,\kappa}_{1,N}$, which, as can 
be verified in appendix {\bf A}, satisfies

\begin{equation}
\left.
X_{\alpha_1,\beta_L}(u_1-v_L)^{\iota,\kappa}_{1,N}
\right|_{e^{u_1}=
\frac{
\beta_L
}
{
\alpha_1
}
e^{v_L}
}
=
0,
\quad
{\rm unless}\ \iota=1,\ \kappa=N 
\end{equation}
Hence, setting $e^{u_1}= \frac{\beta_L}{\alpha_1 } e^{v_L}$ in 
$Z^{DA}_{L\times L}$ freezes the lower-left vertex 
to a type $c{+}$, the remainder of the bottom row to type $a_{-}$, 
and the remainder of the left-most column to type $a_{+}$. Equation 
{\bf \ref{recursion2}} follows from these considerations.

\bigskip
\noindent {\bf Property 4.} The DWPF on a $1\times 1$ lattice is 
given by the $c_{+}$ vertex

\begin{equation}
Z^{DA}_{1\times 1}
=
e^{(N-1)(u_1-v_1)}
\prod_{j=1}^{N-1}
\ll
\sqrt{1-\rho^{j-1} \alpha_1^2}
\sqrt{1-\rho^{j-1} \beta_1^2}
\rr
\end{equation}

\noindent which follows from the definition of the DWBC and the 
weights.

\begin{lemma}
{\rm The above four properties determine the $N$--DA DWPF, 
$N \in \{2, 3, 4\}$, uniquely.

\bigskip

\noindent {\it Proof.} Write $Z^{DA}_{L\times L}$ for the actual 
DWPF, and assume there exists some other 
$\Pi^{DA}_{L\times L}$ which satisfies all of the preceding 
four properties. By Property {\bf 4}, we have 
$Z^{DA}_{1\times 1} = \Pi^{DA}_{1\times 1}$, which is the 
basis for induction. Fix an integer $n \geq 2$. From Properties 
{\bf 1} and {\bf 2}, $\Pi^{DA}_{n \times n}$ must be equal to 
$Z^{DA}_{n \times n}$, up to a multiplicative term, $\mathcal{C}$, 
that does not depend on $e^{u_1}$.

From Property {\bf 3} and the inductive assumption 
$Z^{DA}_{(n-1)\times (n-1)}=\Pi^{DA}_{(n-1)\times (n-1)}$, we 
find that the multiplicative constant $\mathcal{C}=1$. Hence, 
$Z^{DA}_{n\times n}=\Pi^{DA}_{n\times n}$, proving the uniqueness 
claim by induction.

}
\end{lemma}

\subsection{Evaluation of the $N$--DA DWPF, $N \in \{2, 3, 4\}$}

We postulate an expression for $Z^{DA}_{L \times L}$, then show that 
it satisfies the four properties of the previous section.

\bigskip

\noindent{\bf Lemma 2.} 

\bigskip 

\begin{boxedminipage}[l]{12cm}
\begin{eqnarray}
\quad
Z^{DA}_{L\times L}
=
\prod_{j=1}^{L}
\ll
e^{(N-1)j(u_j-v_j)}
\prod_{k=1}^{N-1}
\sqrt{1-\rho^{k-1} \alpha_j^2}
\sqrt{1-\rho^{k-1} \beta_j^2}
\rr
\times
\nonumber
\\
\phantom{\times}
\prod_{1 \leq i < j \leq L}
\prod_{k=1}^{N-1}
\ll 
1 - \rho^{k-1} \alpha_i \alpha_j e^{u_i-u_j} 
\rr 
\ll
1 - \rho^{k-1} \beta_j \beta_i e^{v_j-v_i}
\rr
\label{productform-DA}
\end{eqnarray}
\end{boxedminipage}

\begin{proof}
By inspection, the product expression in Equation 
{\bf \ref{productform-DA}} satisfies Property 
{\bf 1} and {\bf 4}. It contains a factor of 
$
\prod_{j=2}^{L} 
\prod_{k=1}^{N-1} 
\ll
1 - \rho^{k-1} \alpha_1 \alpha_j e^{u_1-u_j}
\rr
$,
which means it possesses the $(L-1)(N-1)$ zeros required by 
Property {\bf 2}. Finally, working directly from Equation 
{\bf \ref{productform-DA}}, we obtain

\begin{equation}
\begin{split}
& Z^{DA}_{L \times L} = 
e^{-L(N-1)v_L}
\prod_{k=1}^{L}
\ll
e^{(N-1)u_k}
\rr
\prod_{j=1}^{N-1}
\ll
\sqrt{1-\rho^{j-1}\alpha_1^2}
\sqrt{1-\rho^{j-1} \beta_L^2}
\rr
\times \\
\label{recursion3}
& \prod_{k=1}^{N-1}
  \prod_{i=1}^{L-1}
\ll
1 - \rho^{k-1} \beta_L \beta_i e^{v_L-v_i}
\rr
\prod_{j=2}^{L}
\ll
1- \rho^{k-1} \alpha_1 \alpha_j e^{u_1-u_j}
\rr Z^{DA, (1L)}_{(L-1) \times (L-1)}
\end{split}
\end{equation}

Evaluating Equation {\bf \ref{recursion3}} at the point 
$e^{u_1}= \frac{\beta_L}{\alpha_1} e^{v_L}$, we obtain 

\begin{equation}
\begin{split}
& Z^{DA}_{L \times L}
\left|_{e^{u_1}=
\frac{\beta_L}{\alpha_1} e^{v_L}}
\right.
=
\ll \frac{\beta_L}{\alpha_1} \rr^{N-1}
e^{-(L-1)(N-1)v_L}   \times \\
& \prod_{k=2}^{L}
\ll e^{(N-1)u_k} \rr 
\prod_{j=1}^{N-1}
\ll
\sqrt{1-\rho^{j-1} \alpha_1^2}
\sqrt{1-\rho^{j-1}  \beta_L^2}
\rr
\times \\
& \prod_{k=1}^{N-1}
  \prod_{i=1}^{L-1}
\ll
1 - \rho^{k-1} \beta_L \beta_i e^{v_L-v_i}
\rr
\prod_{j=2}^{L}
\ll
1- \rho^{k-1} 
\beta_L 
\alpha_j 
e^{v_L-u_j}
\rr
Z^{DA, (1L)}_{(L-1) \times (L-1)}
\label{recursion4}
\end{split}
\end{equation}

Rearranging factors in Equation {\bf \ref{recursion4}}, one recovers 
Equation {\bf \ref{recursion2}} in Property {\bf 3}, as required. This 
concludes our proof that the $N$--DA DWPF, $N \in \{2, 3, 4\}$,  
factorizes.

\end{proof}

{\bf Conjecture.} Equation {\bf \ref{productform-DA}} 
is valid for all $N \geq 2 $. This conjecture is based on the fact that 
our results for $N \in \{2, 3, 4\}$ have a uniform dependence on $N$, 
in which only the $c_{+}$ and line-permuting vertices appear. A study 
of these very vertices for a few values of $N > 4$, indicates that they 
have analogous forms and properties, hence our conjecture.  However, 
a proof of this conjecture requires detailed knowledge of all vertices, 
for all $N > 4$, which is beyond the scope of this work.

\section{The $sl(r+1|s+1)$ Perk-Schultz (PS) models}

This section will be brief, as the arguments are the same as for the 
$N$-DA models. The PS models are defined on the same lattice, with 
the same orientations as the $N$--DA models. The differences are in 
the weights, as those of the PS models do not depend on external 
field variables. 

\subsection{Two sets of state variables} Following \cite{tsuboi}, 
we define two sets 
$B_{-}$$ = $$\{1,   $$\ldots, $$s+1  \}$,
$B_{+}$$ = $$\{s+2, $$\ldots, $$r+s+2\}$,
and their union

$$
B =
\{
\underbrace{1,   \dots,   s+1    }_{B_-},
\underbrace{s+2, \dots, r+s+2 = N}_{B_+}
\}
$$

\subsection{The weights} Let $a,b \in B$. The non-vanishing weights 
of the $sl(r+1|s+1)$ Perk-Schultz (PS) models are

\begin{equation*}
R^{a,a}_{a,a}(u) =
\left\{
\begin{array}{cc}
\frac{\sinh \eta(1-u)}{\sinh\eta},  & a \in B_- \\
\frac{\sinh \eta(1+u)}{\sinh\eta},  & a \in B_+
\end{array}
\right. 
\end{equation*}

\begin{equation*}
R^{a,b}_{b,a}(u) =
\left\{
\begin{array}{cc}
-
\frac{\sinh \eta u}{\sinh \eta}, &
a, b \in{B_-}\   
{\rm or}\    
a, b \in{B_+}\ 
\\
\phantom{-}
\frac{\sinh \eta u}{\sinh \eta}, &
{\rm otherwise}
\end{array}
\right.
\end{equation*}

\begin{equation}
R^{a, b}_{a, b}(u) = 
\left\{
\begin{array}{cc} 
e^{+\eta u}, & a < b   \\
e^{-\eta u}, & a > b
\end{array}
\right.
\label{PS-weights}
\end{equation}

\noindent where $\eta$ is a crossing parameter. The labelling of
the vertices follows the same convention as in Figure {\bf 2}.

\subsection{Non-invariance under state variable conjugation.} 
It is clear, by inspection, that the weights of the $sl(r+1|s+1)$ 
PS model are not invariant under conjugation of state variables, 
where $N=r+s+2$.

\subsection{Symmetries and $\{r, s\}$-independence} From Equations 
{\bf \ref{PS-weights}}, it is clear that the PS weights are 
symmetric in $\sigma_{-} \in B_{-}$ and (separately) in the 
$\sigma_{+} \in B_{+}$\footnote{$sl(N)$ Belavin models are 
analogously symmetric, but all state variables take values in 
one set only $\sigma \in \{1, \cdots, N\}$.}. One can also see 
that, choosing any $\sigma_{-} \in B_{-}$ and any $\sigma_{+} 
\in B_{+}$, as domain wall boundary variables, no other state 
variables appear in the domain wall configurations, and one 
obtains a factorized DWPF that is independent of $\{r, s\}$, 
in other words, the same result one obtains in the $sl(1|1)$ 
model\footnote{In $sl(N)$ Belavin models, choosing any two 
distinct state variables to impose DWBC's, no other state 
variables appear in the configurations, and one obtains Izergin's 
determinant expression for the spin-$\frac{1}{2}$ model, which 
corresponds to $N = 2$.}.

\subsection{PS Domain wall boundary conditions} We define the
$sl(r+1|s+1)$ PS DWBC as follows:
The state variables on all bonds on the right and lower boundaries
are maximal, $\sigma = N=r+s+2 \in B_{+}$, and
the state variables on all bonds on the left and upper boundaries
are minimal, $\sigma = 1 \in B_{-}$.

Notice that, using the symmetries of the weights, we could have 
taken {\it any} state variable $\sigma \in B_{-}$ on the left and 
top, and any state variable $\sigma \in B_{+}$ on the right and 
below. The advantage of the above choice is that the labels of the 
$c_{+}$ and line-permuting vertices are precisely the same as those 
in the $N$--DA models. 

\subsection{Properties of the $sl(r+1|s+1)$ PS DWPF}
The $sl(r+1|s+1)$ PS DWPF, $Z^{PS}_{L\times L}$, satisfies the
following four properties, the proofs of which are precisely 
analogous to those of the $N$--DA models. 

\bigskip

\noindent {\bf Property 1.}  Given the PS weights and DWBC's, and 
writing $U_1 = e^{\eta u_1}$, $Z^{PS}_{L\times L}$ has the form 

\begin{equation}
Z^{PS}_{L\times L}\ll \{u\}, \{v\} \rr
=
U_1^{-L+2} p \ll \{u\}, \{v\} \rr
\end{equation}

\noindent where $p$ is a polynomial of degree $(L-1)$ in $U_1^2$.

\noindent {\bf Property 2.} Using the line-permuting vertices,
$\{R^{1,1}_{1,1}, R^{N,N}_{N,N}\}$, and the Yang-Baxter equations, 
it is straightforward to show that

\begin{equation*}
Z^{PS}_{L\times L} \ll \{u\},\{v \}\rr =
\prod^{L}_{j=2}
\frac{R^{1,1}_{1,1}(u_1 - u_j)}{R^{N,N}_{N,N}(u_1 - u_j)}
Z^{PS}_{L\times L} \ll u_2, \dots, u_L, u_1,\{v\}\rr
\end{equation*}

\noindent which gives the $(L-1)$ zeros of $p$. 

\noindent {\bf Property 3.} Setting $u_1=v_L$, we freeze the lower 
left-hand corner to an $R^{1,N}_{1,N}$, and obtain the recursion 
relation 

\begin{equation}
\begin{split}
& Z^{PS}_{L\times L}|_{u_1=v_L} = \\
& R^{1,N}_{1,N}(0)
\ll
\prod^{L - 1}_{j = 1} R^{1,1}_{1,1}(v_L - v_j)
\rr
\ll
\prod^{L}_{j = 2} R^{N,N}_{N,N}(u_j - v_L)
\rr
Z^{PS, (1L)}_{(L-1)\times (L-1)}
\end{split}
\end{equation}

\noindent {\bf Property 4.} The initial condition is given by 
the $c_{+}$ vertex

$$ 
Z^{PS}_{1\times 1}(u_1,v_1)=  R^{1,N}_{1,N}(u_1-v_1)
$$

\noindent {\bf Lemma 3.}
The above four properties determine the PS DWPF uniquely.

\bigskip

\noindent {\it Proof.} The proof is identical to that of Lemma
{\bf 1}.

\subsection{Evaluation of the PS DWPF} 

We postulate an expression for $Z^{PS}_{L \times L}$, then show 
that it satisfies the four properties of the previous section.

\bigskip

\noindent {\bf Lemma 4.} 
\bigskip

\begin{boxedminipage}[l]{12cm}
\begin{equation}
\begin{split}
Z^{PS}_{L\times L} &\ll \{u\},\{v \}\rr = \\
&\ll
\prod^{L}_{k=1} R^{1,N}_{1,N} (u_k - v_k)
\rr
\ll
\prod_{1 \le i < j \le L}
R^{1,1}_{1,1}(u_i - u_j)
R^{1,1}_{1,1}(v_j - v_i)
\rr
\label{PS-product}
\end{split}
\end{equation}
\end{boxedminipage}

\bigskip

\noindent {\sl Proof.} The proof is identical to that of Lemma {\bf 2} 
in the case of the $N$-DA models, as one can show that the expression 
in Equation {\bf \ref{PS-product}} obeys the required four 
properties\footnote{In \cite{zz}, a determinant expression for the DWPF 
of the $sl(1|1)$ PS model was obtained. It is straightforward to show 
that it evaluates to the above product expression.}.

\section{Remarks}

Domain wall partition functions are important in physics because 
of their role in the algebraic Bethe ansatz approach to correlation 
functions \cite{korepin-book}, and in combinatorics, because in the 
special case of the spin-$\frac{1}{2}$ model, they lead to 
counting alternating sign matrices \cite{kuperberg}.

The factorizing DWPF's discussed in this work do not lead to new 
combinatorics: One can define corresponding combinatorial objects, 
but the weights do not lead to 1-counting (they cannot all be 
simultaneously set to 1). The factorization of the DWPF's (probably) 
reflects the fermionic nature of the underlying models and the absence 
of interesting counting is in turn a reflection of that nature.

On the other hand, factorized DWPF's are easier to handle than 
determinants and one expects (from experience with the
spin-$\frac{1}{2}$ model at the free fermion point) that computing 
the corresponding correlation functions will be easier than in 
generic models. 

The main point of this work is to observe that, if the weights 
of the line-permuting vertices have different zeros, then the 
corresponding DWPF's factorize. The models discussed in this 
paper offer trigonometric vertex examples of this observation. 
It is possible that all such models are either fermionic (with 
or without external fields) as in the $N$--DA models, or contain 
fermions, as in the $sl(r+1|s+1)$ PS models.

\section*{Appendix A}

In the following, $w=u-v$, $x = e^{w}$ and 
$\rho = e^{\frac{2 \pi i n}{N}}$, where $\{n, N\}$ are co-prime.
There is no crossing parameter because these models are external
field deformations of spin-$\frac{N-1}{2}$ models with a crossing
parameter set to the free fermion value.

\subsection*{N=2} (6 vertices)

\begin{equation*}
\begin{split}
&X_{\alpha,\beta}(w)^{1,1}_{1,1}
=
(1 - \alpha \beta x)
&X_{\alpha,\beta}(w)^{1,2}_{1,2}
=
x \sqrt{(1 - \alpha^2)(1 - \beta^2)} \\
&X_{\alpha,\beta}(w)^{1,2}_{2,1}
=
(\alpha- \beta x)
&X_{\alpha,\beta}(w)^{2,1}_{1,2}
=
(\beta- \alpha x) \\
&X_{\alpha,\beta}(w)^{2,1}_{2,1}
=
\sqrt{(1 - \alpha^2)(1 - \beta^2)}
&X_{\alpha,\beta}(w)^{2,2}_{2,2}
=
(x - \alpha \beta)
\end{split}
\end{equation*}


\subsection*{N=3} (19 vertices)

\begin{equation*}
X_{\alpha,\beta}(w)^{1,1}_{1,1} 
=  
(1 - \alpha \beta x )(1 - \alpha \beta \rho x)
\end{equation*}
\begin{equation*}
X_{\alpha,\beta}(w)^{1,2}_{1,2} 
=  
x\sqrt{(1-\alpha^2)(1-\beta^2)}(1-\alpha \beta \rho x)   
\end{equation*}
\begin{equation*}
X_{\alpha,\beta}(w)^{1,2}_{2,1} 
=  
(\alpha- \beta x )(1- \alpha \beta \rho x)
\end{equation*}
\begin{equation*}
X_{\alpha,\beta}(w)^{1,3}_{1,3} 
=  
x^2\sqrt{(1-\alpha^2)(1-\alpha^2 \rho)(1-\beta^2)(1-\beta^2
\rho)}   
\end{equation*}
\begin{equation*}
X_{\alpha,\beta}(w)^{1,3}_{2,2} 
=  
\sqrt{(1-\alpha^2)(1-\beta^2 \rho)} 
\frac{\sqrt{1-\rho^2}}{\sqrt{1-\rho^{\phantom{2}}}}x (\alpha-\beta x)
\end{equation*}
\begin{equation*}
X_{\alpha,\beta}(w)^{1,3}_{3,1} 
=  
(\alpha- \beta x)(\alpha- \beta \rho x)
\end{equation*}
\begin{equation*}
X_{\alpha,\beta}(w)^{2,1}_{1,2} 
=  
(\beta- \alpha x)(1- \alpha \beta \rho x)
\end{equation*}
\begin{equation*}
X_{\alpha,\beta}(w)^{2,1}_{2,1} 
=  
\sqrt{(1-\alpha^2)(1-\beta^2)}(1-\alpha \beta \rho x)
\end{equation*}
\begin{equation*}
X_{\alpha,\beta}(w)^{2,2}_{1,3} 
= 
\sqrt{(1-\alpha^2 \rho)(1-\beta^2)}
\frac{\sqrt{1-\rho^2}}{\sqrt{1-\rho^{\phantom{2}}}} 
x (\beta-\alpha x) 
\end{equation*}
\begin{equation*}
X_{\alpha,\beta}(w)^{2,2}_{2,2} 
= 
(1-\alpha^2)(1-\beta^2 \rho)x
-
(\beta - \alpha x)(\beta x - \alpha \rho) 
\end{equation*}
\begin{equation*}
X_{\alpha,\beta}(w)^{2,2}_{3,1} 
= 
\sqrt{(1-\alpha^2 )(1-\beta^2 \rho)}
\frac{\sqrt{1-\rho^2}}{\sqrt{1-\rho^{\phantom{2}}}} 
(\alpha-\beta x) 
\end{equation*}
\begin{equation*}
X_{\alpha,\beta}(w)^{2,3}_{2,3} 
= 
x (x-\alpha \beta) \sqrt{(1-\alpha^2 \rho)(1-\beta^2 \rho)} 
\end{equation*}
\begin{equation*}
X_{\alpha,\beta}(w)^{2,3}_{3,2} 
= 
(1+\rho)(\alpha-\beta x)(x-\alpha \beta) 
\end{equation*}
\begin{equation*}
X_{\alpha,\beta}(w)^{3,1}_{1,3} 
= (\beta-\alpha x)(\beta- \alpha \rho x) 
\end{equation*}
\begin{equation*}
X_{\alpha,\beta}(w)^{3,1}_{2,2} 
= 
\sqrt{(1-\beta^2)(1-\alpha^2 \rho)} 
\frac{\sqrt{1-\rho^2}}{\sqrt{1-\rho^{\phantom{2}}}}  
(\beta-\alpha x)
\end{equation*}
\begin{equation*}
X_{\alpha,\beta}(w)^{3,1}_{3,1} 
= 
\sqrt{(1-\alpha^2)(1-\alpha^2 \rho)
(1-\beta^2)(1-\beta^2 \rho)} 
\end{equation*}
\begin{equation*}
X_{\alpha,\beta}(w)^{3,2}_{2,3} 
= 
(1+\rho)(\beta-x \alpha)(x-\alpha \beta) 
\end{equation*}
\begin{equation*}
X_{\alpha,\beta}(w)^{3,2}_{3,2} 
= 
\sqrt{(1-\alpha^2 \rho)(1-\beta^2 \rho)}
(x-\alpha \beta) 
\end{equation*}
\begin{equation*}
X_{\alpha,\beta}(w)^{3,3}_{3,3} 
= (x-\alpha \beta)(x-\alpha \beta \rho) 
\end{equation*}


\subsection*{N=4} (44 vertices)

\begin{equation*}
X_{\alpha,\beta}(w)^{1,1}_{1,1} 
=  
(1- \alpha \beta          x)
(1- \alpha \beta \rho   x)
(1- \alpha \beta \rho^2 x)
\end{equation*}
\begin{equation*}
X_{\alpha, \beta}(w)^{1,2}_{1,2} 
=  
x
\sqrt{(1-\alpha^2)(1-\beta^2)}(1-\alpha \beta \rho x)
(1-\alpha \beta \rho^2 x)   
\end{equation*}
\begin{equation*}
X_{\alpha,\beta}(w)^{1,2}_{2,1} 
=  
(   \alpha - \beta        x)
(1- \alpha \beta \rho   x)
(1- \alpha \beta \rho^2 x)
\end{equation*}
\begin{equation*}
X_{\alpha,\beta}(w)^{1,3}_{1,3} 
=  
x^2
\sqrt{(1-\alpha^2)(1-\alpha^2 \rho) (1-\beta^2) (1-\beta^2 \rho)}
(1-\alpha \beta \rho^2 x)   
\end{equation*}
\begin{equation*}
X_{\alpha,\beta}(w)^{1,3}_{2,2} 
=  
\sqrt{(1-\alpha^2)(1-\beta^2 \rho)} 
\frac{\sqrt{1-\rho^2}}{\sqrt{1-\rho^{\phantom{2}}}}x 
(\alpha-\beta x)
(1-\alpha \beta \rho^2 x)
\end{equation*}
\begin{equation*}
X_{\alpha,\beta}(w)^{1,3}_{3,1} 
=  
(\alpha- \beta x)(\alpha- \beta \rho x)(1-\alpha \beta
\rho^2 x)
\end{equation*}
\begin{equation*}
X_{\alpha,\beta}(w)^{1,4}_{1,4} 
=  
x^3 \sqrt{(1-\alpha^2)(1-\alpha^2 \rho)(1-\alpha^2
\rho^2)}\sqrt{(1-\beta^2)(1-\beta^2 \rho)(1-\beta^2 \rho^2)}
\end{equation*}
\begin{equation*}
X_{\alpha,\beta}(w)^{1,4}_{2,3} 
=  
\sqrt{(1-\alpha^2)(1-\alpha^2 \rho)(1-\beta^2)(1-\beta^2 \rho)} 
\frac{\sqrt{1-\rho^3}}{\sqrt{1-\rho^{\phantom{2}}}}
x^{2}(\alpha-\beta x)
\end{equation*}
\begin{equation*}
X_{\alpha,\beta}(w)^{1,4}_{3,2} 
=  
\sqrt{(1-\alpha^2)(1-\beta^2 \rho^2)} 
\frac{\sqrt{1-\rho^3}}{\sqrt{1-\rho^{\phantom{2}}}} x 
(\alpha-\beta x)(\alpha - \beta \rho x)
\end{equation*}
\begin{equation*}
X_{\alpha,\beta}(w)^{1,4}_{4,1} 
= 
(\alpha - \beta          x)
(\alpha - \beta \rho   x)
(\alpha - \beta \rho^2 x)
\end{equation*}
\begin{equation*}
X_{\alpha,\beta}(w)^{2,1}_{1,2} 
=  
(\beta- \alpha x)
(1 - \alpha \beta \rho   x)
(1 - \alpha \beta \rho^2 x)
\end{equation*}
\begin{equation*}
X_{\alpha,\beta}(w)^{2,1}_{2,1} 
=  
\sqrt{(1-\alpha^2)(1-\beta^2)}
(1-\alpha \beta \rho x)
(1-\alpha \beta \rho^2 x)
\end{equation*}
\begin{equation*}
X_{\alpha,\beta}(w)^{2,2}_{1,3} 
= 
\sqrt{(1-\alpha^2 \rho)(1-\beta^2)}
\frac{\sqrt{1-\rho^2}}{\sqrt{1-\rho^{\phantom{2}}}} x 
(\beta-\alpha x)(1-\alpha \beta \rho^2 x) 
\end{equation*}
\begin{equation*}
X_{\alpha,\beta}(w)^{2,2}_{2,2} 
= 
((1-\alpha^2)
(1-\beta^2 \rho)x-
(\beta - \alpha x)
(\beta x - \alpha \rho))
(1-\alpha \beta \rho^2 x) 
\end{equation*}
\begin{equation*}
X_{\alpha,\beta}(w)^{2,2}_{3,1} 
= 
\sqrt{(1-\alpha^2 )(1-\beta^2 \rho)}
\frac{\sqrt{1-\rho^2}}{\sqrt{1-\rho^{\phantom{2}}}}
(\alpha-\beta x)(1-\alpha \beta \rho^2 x) 
\end{equation*}
\begin{equation*}
X_{\alpha,\beta}(w)^{2,3}_{1,4} 
= 
\sqrt{(1-\alpha^2)(1-\alpha^2 \rho)(1-\beta^2)(1-\beta^2 \rho)}
\frac{\sqrt{1-\rho^3}}{\sqrt{1-\rho^{\phantom{2}}}}
(\alpha - \beta                 x)
(1      - \alpha \beta \rho^2 x)   
\end{equation*}
\begin{equation*}
X_{\alpha,\beta}(w)^{2,3}_{2,3} 
=
\sqrt{(1-\alpha^2 \rho)(1-\beta^2 \rho)}
((1-\beta^2)(1-\alpha^2 \rho^2)-(1+\rho)x(\alpha x -\beta \rho)
(\alpha - \beta x)) 
\end{equation*}
\begin{equation*}
X_{\alpha,\beta}(w)^{2,3}_{3,2} 
= 
(\alpha-\beta x)
((1-\alpha^2\beta^2)(1-\rho^3)x - \rho(\alpha x - \beta \rho)
(\alpha-\beta x)) 
\end{equation*}
\begin{equation*}
X_{\alpha,\beta}(w)^{2,3}_{4,1} 
= 
\frac{\sqrt{1-\rho^3}}{\sqrt{1-\rho^{\phantom{2}}}}
\sqrt{(1-\alpha^2)(1-\beta^2 \rho^2)}
(\alpha-\beta x)(\alpha-\beta \rho x) 
\end{equation*}
\begin{equation*}
X_{\alpha,\beta}(w)^{2,4}_{2,4} 
= 
x^{2}\sqrt{1-\alpha^2 \rho} 
\sqrt{(1-\alpha^2 \rho^2)(1-\beta^2 \rho)(1-\beta^2 \rho^2)}
(x-\alpha \beta)
\end{equation*}
\begin{equation*}
X_{\alpha,\beta}(w)^{2,4}_{3,3}
=
x
\frac{
\sqrt{(1-\rho^2)(1-\rho^3)}
}{
1-\rho}
\sqrt{(1-\alpha^2 \rho)(1-\beta^2 \rho^2)}
(x - \alpha \beta)
(\alpha - \beta x)
\end{equation*}
\begin{equation*}
X_{\alpha,\beta}(w)^{2,4}_{4,2} 
=
\frac{(1-\rho^3)
      (x-\alpha \beta)
      (\alpha-\beta x)
      (\alpha-\beta \rho x)}{1-\rho^{\phantom{2}}}
\end{equation*}
\begin{equation*}
X_{\alpha,\beta}(w)^{3,1}_{1,3} 
= 
(\beta-\alpha x)
(\beta- \alpha \rho x) 
(1-\alpha \beta \rho^2 x)
\end{equation*}
\begin{equation*}
X_{\alpha,\beta}(w)^{3,1}_{2,2} 
= 
\sqrt{(1-\beta^2)(1-\alpha^2 \rho)} 
\frac{\sqrt{1-\rho^2}}{\sqrt{1-\rho^{\phantom{2}}}}  
(\beta-\alpha x)(1-\alpha \beta \rho^2 x)
\end{equation*}
\begin{equation*}
X_{\alpha,\beta}(w)^{3,1}_{3,1} 
= 
\sqrt{(1-\alpha^2)(1-\alpha^2 \rho)
(1-\beta^2)
(1-\beta^2 \rho)}
(1-\alpha \beta \rho^2 x) 
\end{equation*}
\begin{equation*}
X_{\alpha,\beta}(w)^{3,2}_{1,4} 
= 
\sqrt{(1-\alpha^2 \rho^2)(1-\beta^2)} 
\frac{\sqrt{1-\rho^3}}{\sqrt{1-\rho^{\phantom{2}}}}x 
(\beta-\alpha x)(\beta-\alpha \rho x) 
\end{equation*}
\begin{equation*}
X_{\alpha,\beta}(w)^{3,2}_{2,3} 
= 
(\beta- \alpha x)
((1-\alpha^2 \beta^2)(1-\rho^3)x
-
\rho(\beta-\alpha x )
(\beta x-\alpha \rho )) 
\end{equation*}
\begin{equation*}
X_{\alpha,\beta}(w)^{3,2}_{3,2} 
= 
\sqrt{(1-\alpha^2 \rho)(1-\beta^2 \rho)}
((1-\alpha^2)(1-\beta^2 \rho^2)x
-
(1+\rho)(\beta-\alpha x)(\beta x-\alpha \rho)) 
\end{equation*}

\begin{equation*}
X_{\alpha,\beta}(w)^{3,2}_{4,1} 
=  
\sqrt{(1-\alpha^2)(1-\alpha^2 \rho)
(1-\beta^2 \rho)(1-\beta^2 \rho^2)}
\frac{\sqrt{1-\rho^3}}{\sqrt{1-\rho^{\phantom{2}}}} 
(\alpha-\beta x)
\end{equation*}
\begin{equation*}
X_{\alpha,\beta}(w)^{3,3}_{2,4} 
=  
x
\frac{\sqrt{(1-\rho^2)(1-\rho^3)}}{1-\rho}
\sqrt{(1-\alpha^2 \rho^2)(1-\beta^2 \rho)}
(x-\alpha \beta)(\alpha-\beta x)
\end{equation*}
\begin{equation*}
X_{\alpha,\beta}(w)^{3,3}_{3,3} 
=  
((1-\alpha^2 \rho)
(1-\beta^2 \rho^2)x-(1+\rho+\rho^2)
(\beta-\alpha x)(\beta x - \alpha \rho))
(x - \alpha \beta)
\end{equation*}
\begin{equation*}
X_{\alpha,\beta}(w)^{3,3}_{4,2} 
=  
\sqrt{(1-\alpha^2 \rho)(1-\beta^2 \rho)}
\frac{\sqrt{(1-\rho^2)(1-\rho^3)}}{1-\rho}
(x-\alpha \beta)(\alpha-\beta x)
\end{equation*}
\begin{equation*}
X_{\alpha,\beta}(w)^{3,4}_{3,4} 
= 
x 
\sqrt{1-\alpha^2 \rho^2}
\sqrt{1-\beta^2 \rho^2}
(x-\alpha \beta)
(x-\alpha \beta \rho)
\end{equation*}
\begin{equation*}
X_{\alpha,\beta}(w)^{3,4}_{4,3} 
=  
\frac{1-\rho^3}{1-\rho}(x-\alpha \beta)
(x-\alpha \beta \rho)(\alpha-\beta x)
\end{equation*}
\begin{equation*}
X_{\alpha,\beta}(w)^{4,1}_{1,4} 
=  
(\beta-\alpha x)
(\beta - \alpha \rho   x)
(\beta - \alpha \rho^2 x)
\end{equation*}
\begin{equation*}
X_{\alpha,\beta}(w)^{4,1}_{2,3} 
=  
\sqrt{(1-\alpha^2 \rho^2)(1-\beta^2)}
\frac{\sqrt{1-\rho^3}}{\sqrt{1-\rho^{\phantom{2}}}}
(\beta-\alpha x)
(\alpha-\beta x)
\end{equation*}
\begin{equation*}
X_{\alpha,\beta}(w)^{4,1}_{3,2} 
=  
\sqrt{(1-\alpha^2)(1-\alpha^2 \rho)(1-\beta^2 \rho)
(1-\beta^2 \rho^2)}
\frac{\sqrt{1-\rho^3}}{\sqrt{1-\rho^{\phantom{2}}}} 
(\beta-\alpha x) 
\end{equation*}
\begin{equation*}
X_{\alpha,\beta}(w)^{4,1}_{4,1} 
=  
\sqrt{(1-\alpha^2)(1-\alpha^2 \rho)
(1-\alpha^2 \rho^2)}
\sqrt{(1-\beta^2)(1-\beta^2 \rho)(1-\beta^2 \rho^2)} 
\end{equation*}
\begin{equation*}
X_{\alpha,\beta}(w)^{4,2}_{2,4} 
=  
\frac{1-\rho^3}{1-\rho^{\phantom{2}}}
(x-\alpha \beta)
(\beta-\alpha x)
(\beta-\alpha \rho x) 
\end{equation*}
\begin{equation*}
X_{\alpha,\beta}(w)^{4,2}_{3,3} 
=  
\sqrt{(1-\alpha^2 \rho^2)(1-\beta^2 \rho)}
\frac{\sqrt{(1-\rho^2)(1-\rho^3)}}{1-\rho}(x-\alpha \beta)
(\beta-\alpha x) 
\end{equation*}
\begin{equation*}
X_{\alpha,\beta}(w)^{4,2}_{4,2} 
=  
\sqrt{(1-\alpha^2)(1-\alpha^2 \rho)
(1-\beta^2 \rho)
(1-\beta^2 \rho^2)}
(x-\alpha \beta) 
\end{equation*}
\begin{equation*}
X_{\alpha,\beta}(w)^{4,3}_{3,4} 
=  
\frac{1-\rho^3}{1-\rho^{\phantom{2}}}
(x-\alpha \beta)
(x - \alpha \beta \rho)
(\beta-\alpha x) 
\end{equation*}
\begin{equation*}
X_{\alpha,\beta}(w)^{4,3}_{4,3} 
=  
\sqrt{(1-\alpha^2 \rho^2)(1-\beta^2 \rho^2)}
(x - \alpha \beta)
(x - \alpha \beta \rho)
\end{equation*}
\begin{equation*}
X_{\alpha,\beta}(w)^{4,4}_{4,4} 
=  
(x-\alpha \beta)
(x - \alpha \beta \rho)
(x  -\alpha \beta \rho^2)
\end{equation*}

\section*{Acknowledgements}
We thank A Caradoc for collaboration on \cite{4-authors}, the results 
of which motivated this work, Profs T~Deguchi and N~Kitanine for 
discussions on \cite{da1} and related topics. OF would like to thank 
Profs P~Bouwknegt, T~Guttmann 
and T~Miwa for gracious hospitality at ANU, MASCOS and Kyoto University, 
respectively, while this work was in progress. MW and MZ are supported 
by an Australian Postgraduate Award (APA).

\end{document}